\documentclass[preprint]{ptephy_v1}%%%%%% to generate preprint number with ptep logo
\preprintnumber{XXXX-XXXX} %%% %%% Insert preprint number here
\usepackage{hyperref}
\pdfoutput=1
\begin{document}

\title{Existence of Mexican-hat dispersion and symmetry group of a layer}

%%%% To generate auto affiliation numbers please use \author{}\affil{} command

\author{Vladimir Damljanovi\'c}
\affil{Institute of Physics Belgrade, University of Belgrade, Pregrevica 118, 11000 Belgrade, Serbia \email{damlja@ipb.ac.rs}}

%%%%\author{Insert second author name here}
%%%%\affil{Insert second author address here}

%%%%\author{Insert third author name here}
%%%%\author[3]{Insert fourth author name here} %%% Use optional bracket [3] to change the respective address
%%%%\affil{Insert third author address here}

%%%%\author{Insert last author name here\thanks{These authors contributed equally to this work}}
%%%%\affil{Insert last author address here}

%%% To include the collaborator name... Please use the command "\collaborator"
%%% For example: \collaborator{ATLAS Collaboration}

\begin{abstract}%
Increased interest in physics of graphene and other two-dimensional materials boosted investigations of band structure near nodal points and lines. In contrast, group theoretical explanation of simple bands (that do not touch other bands), is sporadically present in the literature. This paper presents electronic dispersions up to forth order in momentum, near Brillouin zone (BZ) high symmetry points of all eighty layer groups. The method applies to non magnetic materials both with or without spin-orbit coupling. Particular attention is devoted to Mexican-hat dispersion, showing that it can appear only at BZ center of hexagonal layer groups. Presented symmetry adapted Taylor expansion of bands can be used to fit ab-initio or experimental band structures, or for analytical calculation of crystal properties. The results presented here might serve also as a guiding tool for design of new two-dimensional materials.
\end{abstract}

\subjectindex{I90, I97}

\maketitle

\section{Introduction}

Discovery of electric field effect in atomically thin graphene increased investigations on two-dimensional (2D) materials. When placed in a vertical magnetic field topological properties of bands become physically measurable. In that sense, most attention is devoted to various type of band crossings, many of which are imposed by symmetry laws \cite{Rev2D, TopMat2}.

On the other hand, simple bands (SB), \emph{i.e.} parts of band structure that do not contain nodal points or lines, also lead to fascinating physics. Two dimensional materials having Mexican-hat dispersion (MHD) are predicted to be multiferroic due to presence of van Hove singularity of the $1/\sqrt{E}$ - type \cite{Se16}. Furthermore, MHD gives that effective mass on two bands extrema changes sign so that originally repulsive electron-electron interaction becomes attractive, giving quasi-bound electron states \cite{Sa23, Sab23}. In addition, materials with MHD have large figure of merit, which makes them suitable for thermoelectric applications \cite{Nur20, Wa21}. For these reasons it would be useful if a general prescription on how to search materials with MHD would exist, and group theory is a powerful tool in this respect.

In 2D, no publications deal with exhaustive symmetry treatment of SB, to best of author knowledge, although two of them treat dispersions in the vicinity of nodal points and lines \cite{Mi23, Enci2D}. On the other hand, encyclopedia type papers on dispersions in 3D \cite{Enci1, Enci2, Enci3, Enci4, Encikp, Encikp24} cannot be used easily for 2D in many cases. While it is straightforward task in crystals with axis of order three, four or six, going back from AA stacking to original layer can be complicated in orthorhombic crystals. Fortunately, recent upgrade in C2DB base \cite{LgClsf} helps in solving this problem.

In this paper, we have used powerful group-theoretical tools to find dispersions of SB in the vicinity of high-symmetry points (HSPs) in the reciprocal space. We have adapted the Taylor expansion to the underlying symmetry up to polynomial of fourth degree. All HSPs of all eighty layer groups, both without and with spin-orbit coupling (SOC) are covered. The results are applicable to non-magnetic layers where the time-reversal (TRS) alone is symmetry of the system. We have illustrated our results using one-band tight-binding model on a structure having the same symmetry as graphene (which belongs to layer group 80).

\section{Method}

Let $\mathbf{k}_0$ be a HSP in the BZ. For $\mathbf{k}=\mathbf{k}_0+\mathbf{q}$ ($|\mathbf{q}|$ is small), the electronic dispersion $E(\mathbf{q})$ has the following properties: $E(\hat{g}\mathbf{q})=E(\mathbf{q})$ and, if $\mathbf{k}_0$ is time-reversal invariant momentum, $E(-\mathbf{q})=E(\mathbf{q})$. Here $\hat{g}$ is rotation/improper rotation from the point group of the wave vector $\underline{G}_0(\mathbf{k}_0)$ in two-dimensional reciprocal space. In that sense, \emph{e.g.} spatial inversion is represented by $-\hat{\sigma}_0$, while the Pauli matrix $\hat{\sigma}_3$, represents $xz$-(symmorphic or glide) reflection plane. TRS in the reciprocal space acts as spatial inversion $\hat{i}$, so that the effective little group is $\underline{G}_0(\mathbf{k}_0)+\hat{i}\underline{G}_0(\mathbf{k}_0)$, if $\underline{G}_0(\mathbf{k}_0)$ does not already contain $\hat{i}$. The groups obtained from $\underline{G}_0(\mathbf{k}_0)$ by adding $\hat{i}$ are given in Table \ref{tab1}.

\begin{table}[!h]
\caption{Correspondence between group $\underline{G}_0(\mathbf{k}_0)$ and the one with spatial inversion added.}%%%Table caption goes here
\label{tab1}
\centering
\begin{tabular}{|c c|c c|c c|}%%%The number of columns has to be defined here
\hline
$\underline{G}_0(\mathbf{k}_0)$ & $\underline{G}_0(\mathbf{k}_0)\otimes\underline{C}_i$ & $\underline{G}_0(\mathbf{k}_0)$ & $\underline{G}_0(\mathbf{k}_0)\otimes\underline{C}_i$ & $\underline{G}_0(\mathbf{k}_0)$ & $\underline{G}_0(\mathbf{k}_0)\otimes\underline{C}_i$\\ %%%% Table body
\hline
$\underline{C}_1$ & $\underline{C}_i$ & $\underline{C}_4$ & $\underline{C}_{4h}$ & $\underline{D}_{2h}$ & $\underline{D}_{2h}$\\%%%% Table body
$\underline{C}_s$ & $\underline{C}_{2h}$ & $\underline{C}_6$ & $\underline{C}_{6h}$ & $\underline{D}_{4h}$ & $\underline{D}_{4h}$\\
$\underline{C}_2$ & $\underline{C}_{2h}$ & $\underline{D}_2$ & $\underline{D}_{2h}$ & $\underline{D}_{6h}$ & $\underline{D}_{6h}$\\
$\underline{C}_3$ & $\underline{S}_6$ & $\underline{D}_4$ & $\underline{D}_{4h}$ & $\underline{D}_{2d}$ & $\underline{D}_{4h}$\\
$\underline{D}_3$ & $\underline{D}_{3d}$ & $\underline{D}_6$ & $\underline{D}_{6h}$ & $\underline{D}_{3d}$ & $\underline{D}_{3d}$\\
$\underline{C}_{2v}$ & $\underline{D}_{2h}$ & $\underline{C}_{4v}$ & $\underline{D}_{4h}$ & $\underline{S}_4$ & $\underline{C}_{4h}$\\
$\underline{C}_{3v}$ & $\underline{D}_{3d}$ & $\underline{C}_{6v}$ & $\underline{D}_{6h}$ & $\underline{S}_6$ & $\underline{S}_6$\\
$\underline{C}_{3h}$ & $\underline{C}_{6h}$ & $\underline{C}_{2h}$ & $\underline{C}_{2h}$ & $\underline{C}_i$ & $\underline{C}_i$  \\
$\underline{D}_{3h}$ & $\underline{D}_{6h}$ & $\underline{C}_{4h}$ & $\underline{C}_{4h}$ & $\underline{C}_{6h}$ & $\underline{C}_{6h}$ \\
\hline
\end{tabular}
\end{table}%%%End of the table

If $E(\mathbf{q})$ is simple band, one can expand $E(\mathbf{q})$ in Taylor series. The coefficients of the expansion (partial derivatives of $E$) are obtained by Wigner method of group projectors ($\otimes$ denotes the Kronecker product):
\begin{equation}
\hat{P}_n=\frac{1}{|\underline{G}_0(\mathbf{k}_0)|}\sum_{g\in\underline{G}_0(\mathbf{k}_0)}\underbrace{\hat{g}\otimes...\otimes\hat{g}.}_{n} \nonumber
\end{equation}
One acts with the projector $\hat{P}_n$ on trial vector of dimension $2^n$ ($n=1, 2, 3,...$). The trial vector must be general but has to reflect symmetry under permutation of coordinates in partial derivatives. For example trial vector for second order derivatives is $|\mathrm{trial2}>=$\small $(\partial^2E/\partial q_1^2,\partial^2E/\partial q_1\partial q_2,\partial^2E/\partial q_2\partial q_1,\partial^2E/\partial q_2^2)^T$\normalsize $=(u,v,v,w)^T$. We have applied this method for all HSPs of all eighty layer groups to find invariant Taylor polynomials up to fourth order in $q_1, q_2$. The results are shown in the following section. The HSPs of layer groups hosting simple bands are derived by omitting nodal lines and points given in \cite{Mi23, Ja23}.

\section{Results}

\begin{figure}[!h]
\centering\includegraphics[width=5in]{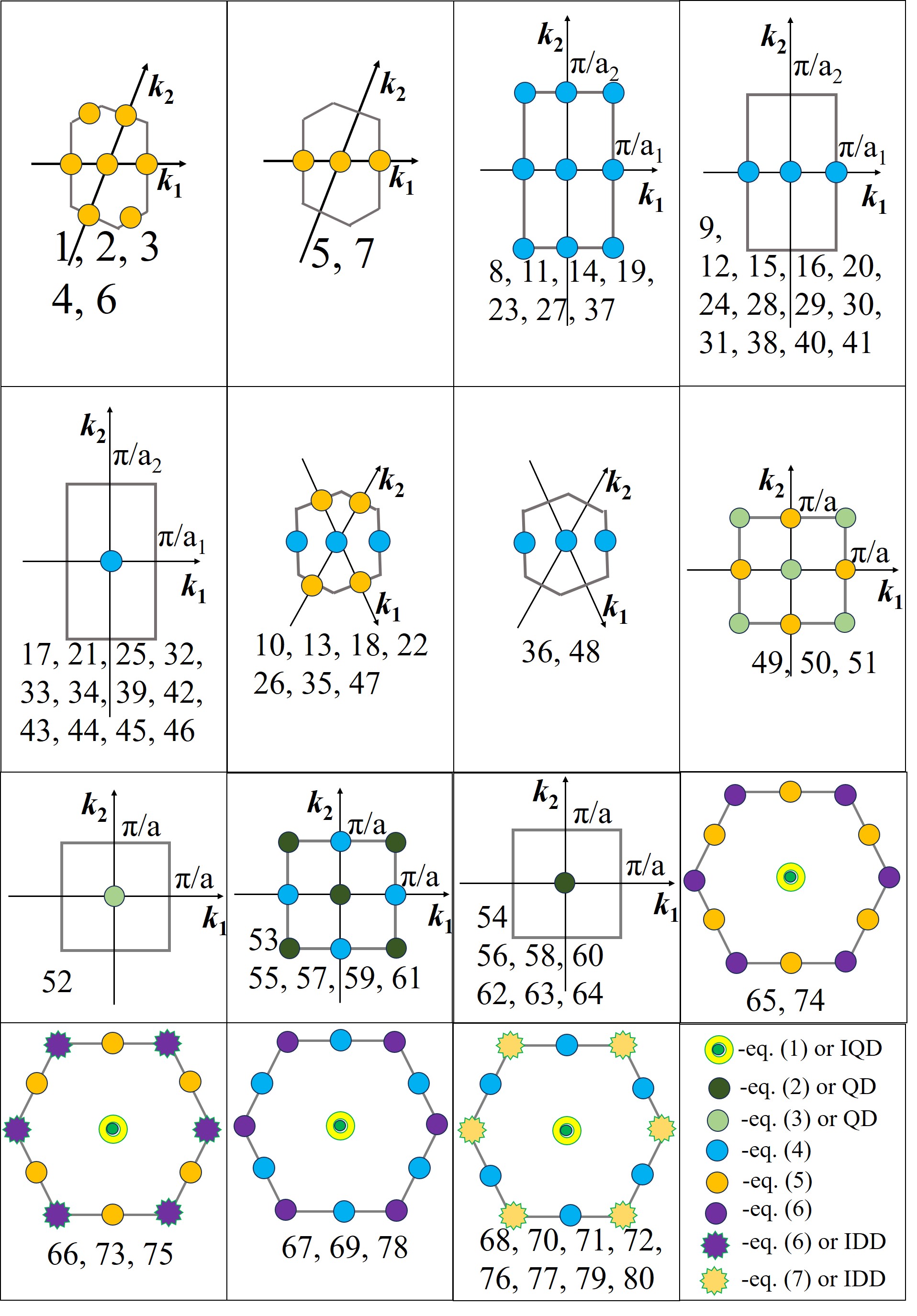}
%%%call your figure name in the place "figurename.eps"
\caption{Positions of simple bands in $k$-space for all gray layer groups without SOC. IDD - isotropic Dirac dispersion: $E_{1,2,3,4}(\mathbf{q})=E_0\pm a|\mathbf{q}|$, QD - quadratic dispersion: $E_{1,2,3,4}(\mathbf{q})=E_0+a\mathbf{q}^2\pm \left\{\left[b_3(q_1^2-q_2^2)+b_1q_1q_2\right]^2+\left[b_4(q_1^2-q_2^2)+b_2q_1q_2\right]^2\right\}^{1/2}$, IQD - isotropic quadratic dispersion: $E_{1,2,3,4}(\mathbf{q})=E_0+(a\pm b)\mathbf{q}^2$.}
\label{fig1}
\end{figure}

\begin{figure}[!h]
\centering\includegraphics[width=5in]{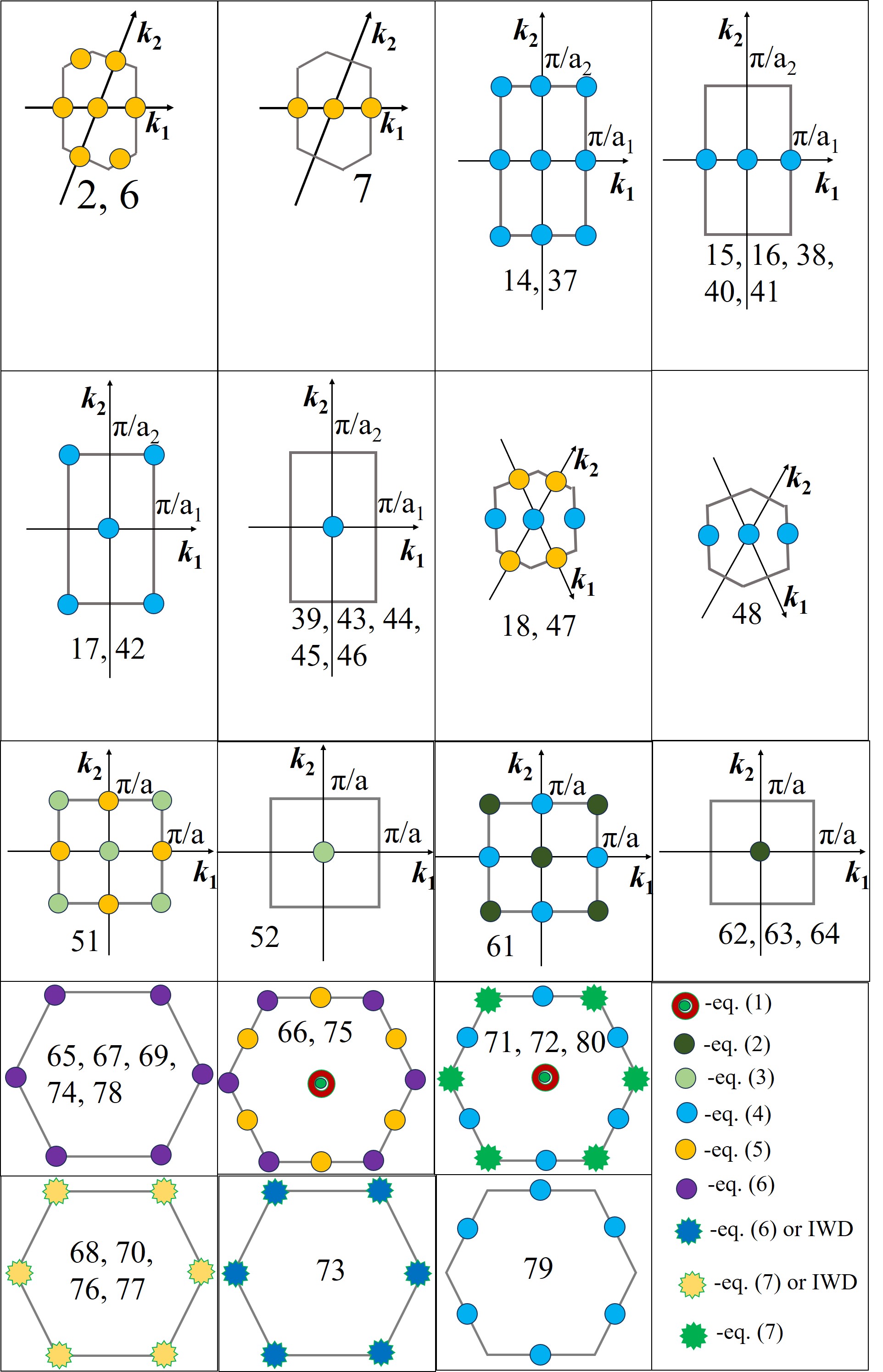}
%%%call your figure name in the place "figurename.eps"
\caption{Positions of simple bands in $k$-space for all gray layer groups with SOC. Superscript $D$ for double groups is omitted. IWD - isotropic Weyl dispersion: $E_{1,2}=E_0\pm a|\mathbf{q}|$.}
\label{fig2}
\end{figure}

We found in total seven dispersion formulas. For little groups $\underline{C}_6$, $\underline{D}_6$, $\underline{C}_{6v}$, $\underline{C}_{6h}$, $\underline{D}_{6h}$, $\underline{S}_6$ and $\underline{D}_{3d}$ one gets the following dipersion ($a$, $b$, $c$, $E_0$ are real parameters):
\begin{equation}
\label{jedan}
E(\mathbf{q})\approx E_0+a\mathbf{q}^2+b\mathbf{q}^4.
\end{equation}
Groups $\underline{D}_{2d}$, $\underline{C}_{4v}$, $\underline{D}_4$ and $\underline{D}_{4h}$ give:
\begin{equation}
\label{dva}
E(\mathbf{q})\approx E_0+a\mathbf{q}^2+b_1(q_1^4+q_2^4)+b_2q_1^2q_2^2,
\end{equation}
groups $\underline{C}_4$, $\underline{C}_{4h}$ and $\underline{S}_4$ give:
\begin{equation}
\label{tri}
E(\mathbf{q})\approx E_0+a\mathbf{q}^2+b_1(q_1^4+q_2^4)+b_2q_1^2q_2^2 +b_3(q_1^3q_2-q_1q_2^3),
\end{equation}
groups $\underline{D}_2$, $\underline{D}_{2h}$, $\underline{C}_{2v}^z$ and $\underline{C}_{2h}^{\mathrm{horizontal}}$ give:
\begin{equation}
\label{cetiri}
E(\mathbf{q})\approx E_0+a_1q_1^2+a_2q_2^2+b_1q_1^4+b_3q_2^4+b_2q_1^2q_2^2,
\end{equation}
groups $\underline{C}_i$, $\underline{C}_2^z$ and $\underline{C}_{2h}^z$ give:
\begin{equation}
\label{pet}
E(\mathbf{q})\approx E_0+\sum_{j,l=1}^2a_{j,l}q_jq_l+\sum_{j,l,n,m=1}^2b_{j,l,n,m}q_jq_lq_nq_m,
\end{equation}
groups $\underline{C}_3$ and $\underline{C}_{3h}$ give:
\begin{equation}
\label{sest}
E(\mathbf{q})\approx E_0+a\mathbf{q}^2+c_1(q_1^3-3q_1q_2^2)+c_2(q_2^3-3q_1^2q_2)+b\mathbf{q}^4,
\end{equation}
while groups $\underline{D}_3$, $\underline{C}_{3v}$ and $\underline{D}_{3h}$ give:
\begin{equation}
\label{sedam}
E(\mathbf{q})\approx E_0+a\mathbf{q}^2+c_1(q_1^3-3q_1q_2^2)+b\mathbf{q}^4.
\end{equation}
With these knowledge we obtain dispersions in all layer gray (single and double) groups. The results are shown in Figure \ref{fig1}, for SOC not included and Figure \ref{fig2}, for the case with SOC. The orientation of the basis vectors in the reciprocal space is the following: for all oblique and rectangular groups having fractional translations parallel to one direction, the vector $\mathbf{k}_1$ is orthogonal to that direction.

\begin{figure}[!h]
\centering\includegraphics[width=3.5in]{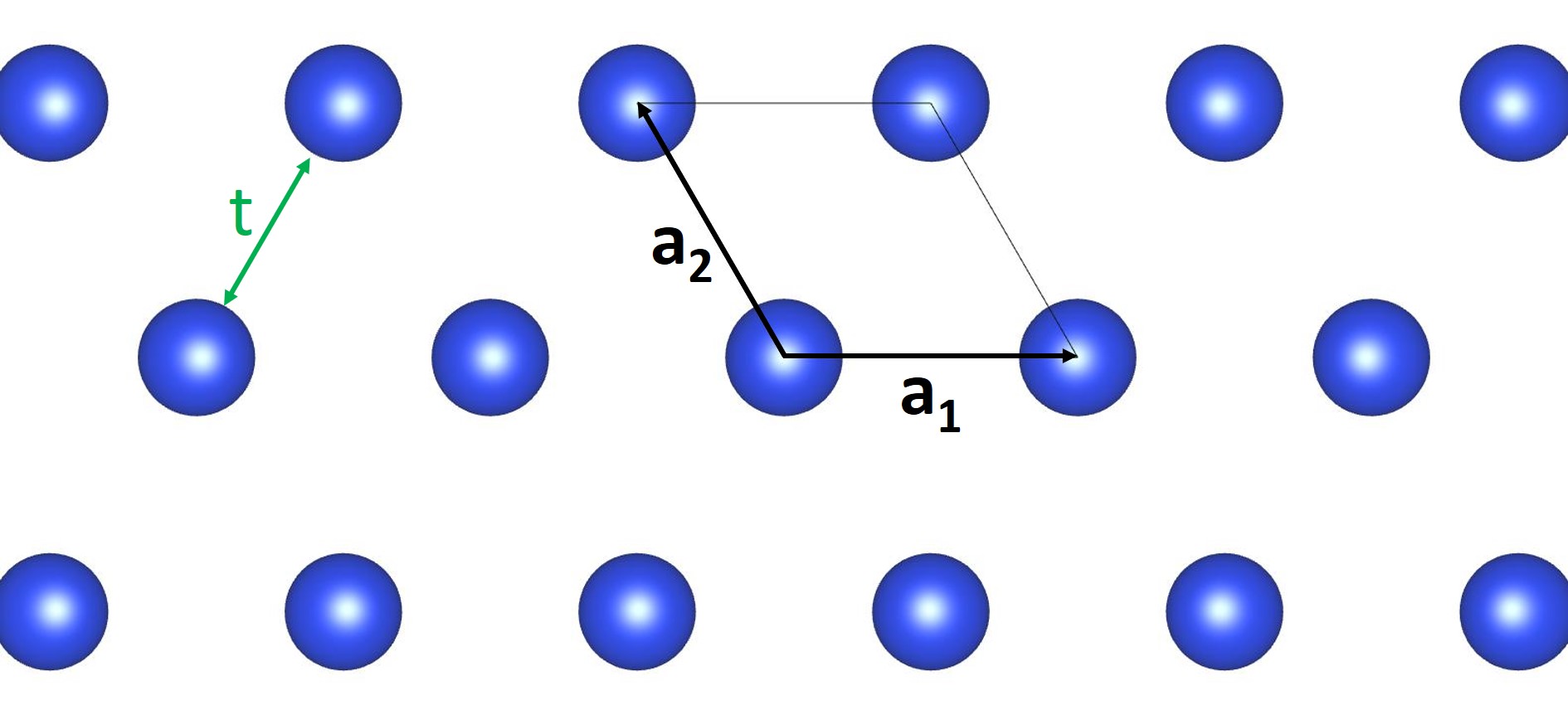}
%%%call your figure name in the place "figurename.eps"
\caption{Lattice structure for the tight-binding model. Black parallelogram denotes primitive unit cell. Hopping parameter \emph{t} between nearest neighbors is indicated in green color.}
\label{fig3}
\end{figure}

As an illustration we consider one site tight-binding model belonging to gray single layer group 80. The crystal structure is shown in the Figure \ref{fig3} (visualization by use of VESTA \cite{vesta}). The Hamiltonian is one dimensional and the band structure is:
\begin{equation}
\label{disp}
E(\mathbf{k})=E_0+2t\left\{\mathrm{cos}(\mathbf{k}\cdot\mathbf{a}_1)+\mathrm{cos}(\mathbf{k}\cdot\mathbf{a}_2)+\mathrm{cos}[\mathbf{k}\cdot(\mathbf{a}_1+\mathbf{a}_2)]\right\}.
\end{equation}
Near BZ center we have $\mathbf{k}=0+\mathbf{q}$ so that the dispersion is:
\begin{equation}
\label{dispg}
E_{\Gamma}(\mathbf{q})\approx E_0+6t-\frac{3}{2}ta^2\mathbf{q}^2+\frac{3}{32}ta^4\mathbf{q}^4.
\end{equation}
For $t>0$ ($t<0$) eq. (\ref{dispg}) presents Mexican Hat dispersion (inverted Mexican Hat dispersion).

Around $\mathrm{M}$-point we have $\mathbf{k}=\mathbf{b}_1/2+\mathbf{q}$ so that:
\begin{equation}
\label{dispm}
E_M(\mathbf{q})\approx E_0-2t+\frac{1}{2}ta^2(3q_1^2-q_2^2)-\frac{1}{96}ta^4(9q_1^4+18q_1^2q_2^2-7q_2^4).
\end{equation}
Around $\mathrm{K}$-point we have ($\mathbf{k}=(\mathbf{b}_1+\mathbf{b}_2)/3+\mathbf{q}$):
\begin{equation}
\label{dispk}
E_K(\mathbf{q})\approx E_0-3t+\frac{3}{4}ta^2\mathbf{q}^2-\frac{\sqrt{3}}{8}ta^3(q_1^3-3q_1q_2^2)-\frac{3}{64}ta^4\mathbf{q}^4.
\end{equation}
The equations (\ref{dispg}), (\ref{dispm}) and (\ref{dispk}) are another confirmations of the theory presented here.

%%%%\section{Enunciations}
%%%% Most of the enunciations like theorem, lemma, corollary, proposition, defintion,
%%%% condition, example, conjecture etc. are defined in the class file.

%%%% If the author wants to add or modify the enunciation style
%%%% they can define in the preamble as shown below.

%%%% \newtheoremstyle{theorem}{6pt}{6pt}{\rm}{}{\sffamily}{ }{ }{}
%%%% \theoremstyle{theorem}
%%%% \newtheorem{theorem}{\sc Theorem}[section]

%%%%\newtheoremstyle{corollary}{6pt}{6pt}{\rm}{}{\sffamily}{ }{ }{}
%%%%\theoremstyle{corollary}
%%%%\newtheorem{corollary}{\sc Corollary}[section]

%%%%\newtheoremstyle{definition}{6pt}{6pt}{\rm}{}{\sffamily}{ }{ }{}
%%%%\theoremstyle{definition}
%%%%\newtheorem{definition}[theorem]{\sc Definition}
%%%%
%%%%\newtheorem{exercise}[theorem]{Exercise}

\section{Discussion}

The signs of parameters in formulas (\ref{jedan}) - (\ref{sedam}) can not be determined using group theory alone, so \emph{e.g.} in formula (\ref{jedan}) it is not guaranteed that $ab<0$, the necessary condition for MHD/inverted MHD. However, in multiband cases, for the lowest lying band, the second order perturbation theory gives $a<0$, which is the first prerequisite for MHD. In addition, Figures \ref{fig1} and \ref{fig2} show that the symmetry forbids appearance of MHD outside of BZ centers of hexagonal groups. Existence of MHD at the $\Gamma$-point is experimentally confirmed in hexagonal $\mathrm{GaSe}$ grown on $\mathrm{GaAs}$ by molecular beam epitaxy \cite{Exp18}. The MHD and high electron mobility are numerically predicted in single layers of $\gamma-\mathrm{SnX}$ ($\mathrm{X}=\mathrm{O, S, Se, Te}$), which are also hexagonal materials, by first principles investigation \cite{Num22}. Similarly, hexagonal lattices of group-VA elements, having symmetry of graphene, are shown analytically to exhibit MHD with magnetic instability and unique thermoelectric properties \cite{Sev17}. First principle calculations show inverted MHD at BZ center in Janus $\gamma-\mathrm{Ge_2XY}$ ($\mathrm{X/Y=S, Se, Te}$) monolayers having symmetry $p3m1$ (layer group 69) \cite{Vu23}. The inverted MHD remains feature in the band structure also when vertical electric field is applied. This effect can be explained having in mind that group $p3m1$ is also a wallpaper group, so it does not contain elements that flip the surface of 2D material. When homogenous electric field is applied perpendicularly to the sample, the symmetry group remains the same.

Formulas other then (\ref{jedan}) are also used in the literature for fitting. Penta-graphene belongs to layer group $p\overline{4}2_1m$ (layer group 58), so that its isogonal group is $\underline{D}_{2d}$ which is non-centrosymmetric \cite{Ji23}. After adding spatial inversion the group becomes $\underline{D}_{4h}$, which requires eq. (\ref{dva}) for fitting bands around BZ center. This exact equation is used in \cite{Ji23} under the name tetragonal MHD.

Sometimes it is necessary to go beyond fourth order invariant polynomials. Hexagonal gallium chalcogenides belong to layer group 78 ($p\overline{6}m2$) and the band structure is fitted with polynomial of order six in \cite{Zo13}. Instead of giving full polynomial of order six, for comparison with the results of \cite{Zo13} here we only calculate the number of real coefficients. Using general formula for symmetrized $n$-th power of a representation \cite{Ly60}, we get: $\left[\Gamma^{\otimes 6}\right]=2A_{1g}+A_{2g}+2E_{2g}$ in the group $\underline{D}_{6h}$. The number of parameters is equal to two, which is in accordance with formula for fitting in \cite{Zo13}. On the other hand, the formula four in \cite{Ry14}, used for fitting of hexagonal $\mathrm{GaS}$ band structure around $\Gamma$-point, can not be justified by symmetry since it contains two coefficients in second order polynomial instead of one. Deviations from completely isotropic dispersion in the vicinity of BZ center are due to higher order terms in the Taylor expansion of electronic energy, as already suggested by \cite{Zo13}.

%%%%\vskip2pc

%%%%\noindent The output for table is:

\section{Conclusion}

In summary we have determined dispersions of simple bands in the vicinity of special points in the reciprocal space for non-magnetic two-dimensional materials without and with spin-orbit coupling. Further research is necessary to clarify cases when TRS alone is broken. This can be the case of magnetic layers, where extension of spin group concepts from \emph{e.g.} quasi one-dimensional materials \cite{Nat1, Nat2, Nat3}, would be appropriate.

Our results are presented in graphical form, not in encyclopedia manner, which might ensure easier usage. The problem of finding new materials is reduced, in this particular case of simple bands, to problem of synthesis of 2D materials with prescribed symmetry. In that sense, available data bases of numerically stable and real materials might be useful \cite{DBase18, DBase21, Exp}. The last point for crystal growers would be the right placement of the Fermi level, which can not be addressed using symmetry alone, although electron filling conditions for symmetry groups of insulating systems can be helpful \cite{ElFil1, ElFil2, ElFil3}.

\section*{Acknowledgment}
Author acknowledges funding by the Ministry of Science, Technological Development and Innovation of the Republic of Serbia provided by the Institute of Physics Belgrade.
\let\doi\relax
%%%%\bibliographystyle{ptephy}
%%%%\bibliography{Ref2404}
%
% once the .bbl file has been generated then place the text in your article.

\vspace{0.2cm}
\noindent
%%%%For references,  note how to include DOI information from examples below. 

%This is added by T. Yoneya (editor-in-chief) on 2020/07/09.

%%%\let\doi\relax

%without this code before the command "\begin{thebibliography}{}" , an error will be %flagged. When the bibliography is provided as separate .bib file, then this code %should be placed above the commands "\bibliographystyle{}" and "\bibliography{}" %inside the main TeX file. 

\begin{thebibliography}{10}

\bibitem{Rev2D}
Xiaolong Feng, Jiaojiao Zhu, Weikang Wu, and Shengyuan~A. Yang, Chinese Physics
  B, {\bf 30}(10), 107304 (nov 2021).

\bibitem{TopMat2}
N.~P. Armitage, E.~J. Mele, and Ashvin Vishwanath, Rev. Mod. Phys., {\bf 90},
  015001 (Jan 2018).

\bibitem{Se16}
L.~Seixas, A.~S. Rodin, A.~Carvalho, and A.~H. Castro~Neto, Phys. Rev. Lett.,
  {\bf 116}, 206803 (May 2016).

\bibitem{Sa23}
Vladimir~A. Sablikov and Aleksei~A. Sukhanov, Physica E: Low-dimensional
  Systems and Nanostructures, {\bf 145}, 115492 (2023).

\bibitem{Sab23}
Vladimir~A. Sablikov and Aleksei~A. Sukhanov, Physics Letters A, {\bf 481},
  129006 (2023).

\bibitem{Nur20}
M~Nurhuda, A~R~T Nugraha, M~Y Hanna, E~Suprayoga, and E~H Hasdeo, Advances in
  Natural Sciences: Nanoscience and Nanotechnology, {\bf 11}(1), 015012 (feb
  2020).

\bibitem{Wa21}
Cong Wang, Zhiyuan Xu, Ke~Xu, and Guoying Gao, Molecules, {\bf 26}(21) (2021).

\bibitem{Mi23}
V~Damljanović and N~Lazić, Journal of Physics A: Mathematical and
  Theoretical, {\bf 56}(21), 215201 (may 2023).

\bibitem{Enci2D}
Zeying Zhang, Weikang Wu, Gui-Bin Liu, Zhi-Ming Yu, Shengyuan~A. Yang, and
  Yugui Yao, Phys. Rev. B, {\bf 107}, 075405 (Feb 2023).

\bibitem{Enci1}
Feng Tang and Xiangang Wan, Phys. Rev. B, {\bf 104}, 085137 (Aug 2021).

\bibitem{Enci2}
Zhi-Ming Yu, Zeying Zhang, Gui-Bin Liu, Weikang Wu, Xiao-Ping Li, Run-Wu Zhang,
  Shengyuan~A. Yang, and Yugui Yao, Science Bulletin, {\bf 67}(4), 375--380
  (2022).

\bibitem{Enci3}
Gui-Bin Liu, Zeying Zhang, Zhi-Ming Yu, Shengyuan~A. Yang, and Yugui Yao, Phys.
  Rev. B, {\bf 105}, 085117 (Feb 2022).

\bibitem{Enci4}
Zeying Zhang, Gui-Bin Liu, Zhi-Ming Yu, Shengyuan~A. Yang, and Yugui Yao, Phys.
  Rev. B, {\bf 105}, 104426 (Mar 2022).

\bibitem{Encikp}
Zeying Zhang, Zhi-Ming Yu, Gui-Bin Liu, Zhenye Li, Shengyuan~A. Yang, and Yugui
  Yao, Computer Physics Communications, {\bf 290}, 108784 (2023).

\bibitem{Encikp24}
João Victor~V. Cassiano, Augusto de~Lelis~Araújo, Paulo E.~Faria Junior, and
  Gerson~J. Ferreira, SciPost Phys. Codebases, page~25 (2024).

\bibitem{LgClsf}
Jingheng Fu, Mikael Kuisma, Ask~Hjorth Larsen, Kohei Shinohara, Atsushi Togo,
  and Kristian~S. Thygesen,
\newblock Layer group classification of two-dimensional materials (2024),
  {{arXiv:2401.16705}}.

\bibitem{Ja23}
Vladimir Damljanović, Progress of Theoretical and Experimental Physics, {\bf
  2023}(4), 043I02 (04 2023),
  {{https://academic.oup.com/ptep/article-pdf/2023/4/043I02/50112917/ptad050.pdf}}.

\bibitem{vesta}
Koichi Momma and Fujio Izumi, Journal of Applied Crystallography, {\bf 44}(6),
  1272--1276 (Dec 2011).

\bibitem{Exp18}
Ming-Wei Chen, HoKwon Kim, Dmitry Ovchinnikov, Agnieszka Kuc, Thomas Heine,
  Olivier Renault, and Andras Kis, npj 2D Materials and Applications, {\bf 2},
  2 (Jan 2018).

\bibitem{Num22}
Vu~V. Tuan, A.~A. Lavrentyev, O.~Y. Khyzhun, Nguyen T.~T. Binh, Nguyen~V. Hieu,
  A.~I. Kartamyshev, and Nguyen~N. Hieu, Phys. Chem. Chem. Phys., {\bf 24},
  29064--29073 (2022).

\bibitem{Sev17}
Hâldun Sevinçli, Nano Letters, {\bf 17}, 2589--2595 (Apr 2017).

\bibitem{Vu23}
Tuan~V Vu, Huynh~V Phuc, Le~C Nhan, A~I Kartamyshev, and Nguyen~N Hieu, Journal
  of Physics D: Applied Physics, {\bf 56}(13), 135302 (mar 2023).

\bibitem{Ji23}
Ningning Jia, Yongting Shi, Zhiheng Lv, Junting Qin, Jiangtao Cai, Xue Jiang,
  Jijun Zhao, and Zhifeng Liu, New Journal of Physics, {\bf 25}(3), 033033 (mar
  2023).

\bibitem{Zo13}
V.~Z\'olyomi, N.~D. Drummond, and V.~I. Fal'ko, Phys. Rev. B, {\bf 87}, 195403
  (May 2013).

\bibitem{Ly60}
G.~Ya. Lyubarskii,
\newblock {\em The Application of Group Theory in Physics},
\newblock  (Pergamon Press, New York, 1960).

\bibitem{Ry14}
Dmitry~V. Rybkovskiy, Alexander~V. Osadchy, and Elena~D. Obraztsova, Phys. Rev.
  B, {\bf 90}, 235302 (Dec 2014).

\bibitem{Nat1}
Nataša Lazić, Marko Milivojević, and Milan Damnjanović, Acta
  Crystallographica Section A, {\bf 69}(6), 611--619 (2013),
  {{https://onlinelibrary.wiley.com/doi/pdf/10.1107/S0108767313022642}}.

\bibitem{Nat2}
Nata\v{s}a Lazi\'c and Milan Damnjanovi\'c, Phys. Rev. B, {\bf 90}, 195447 (Nov
  2014).

\bibitem{Nat3}
Nata\v{s}a Lazi\'c,
\newblock {\em QUASI-CLASSICAL GROUND STATES AND MAGNONS IN MONOPERIODIC SPIN
  SYSTEMS},
\newblock {Ph.D.} thesis, Faculty of Physics, University of Belgrade (October
  2016).

\bibitem{DBase18}
Sten Haastrup, Mikkel Strange, Mohnish Pandey, Thorsten Deilmann, Per~S
  Schmidt, Nicki~F Hinsche, Morten~N Gjerding, Daniele Torelli, Peter~M Larsen,
  Anders~C Riis-Jensen, Jakob Gath, Karsten~W Jacobsen, Jens~Jørgen Mortensen,
  Thomas Olsen, and Kristian~S Thygesen, 2D Materials, {\bf 5}(4), 042002 (sep
  2018).

\bibitem{DBase21}
Morten~Niklas Gjerding, Alireza Taghizadeh, Asbjørn Rasmussen, Sajid Ali,
  Fabian Bertoldo, Thorsten Deilmann, Nikolaj~Rørbæk Knøsgaard, Mads Kruse,
  Ask~Hjorth Larsen, Simone Manti, Thomas~Garm Pedersen, Urko Petralanda,
  Thorbjørn Skovhus, Mark~Kamper Svendsen, Jens~Jørgen Mortensen, Thomas
  Olsen, and Kristian~Sommer Thygesen, 2D Materials, {\bf 8}(4), 044002 (jul
  2021).

\bibitem{Exp}
Nicolas Mounet, Marco Gibertini, Philippe Schwaller, Davide Campi, Andrius
  Merkys, Antimo Marrazzo, Thibault Sohier, Ivano~Eligio Castelli, Andrea
  Cepellotti, Giovanni Pizzi, and Nicola. Marzari, Nature Nanotechnology, {\bf
  13}, 246 -- 252 (Mar 2018).

\bibitem{ElFil1}
Haruki Watanabe, Hoi~Chun Po, Michael~P. Zaletel, and Ashvin Vishwanath, Phys.
  Rev. Lett., {\bf 117}, 096404 (Aug 2016).

\bibitem{ElFil2}
Haruki Watanabe, Hoi~Chun Po, Ashvin Vishwanath, and Michael Zaletel,
  Proceedings of the National Academy of Sciences, {\bf 112}(47), 14551--14556
  (2015),  {{https://www.pnas.org/doi/pdf/10.1073/pnas.1514665112}}.

\bibitem{ElFil3}
Benjamin~J. Wieder and C.~L. Kane, Phys. Rev. B, {\bf 94}, 155108 (Oct 2016).

\end{thebibliography}

%%%%\begin{thebibliography}{9}

%%%%\bibitem{1}
%%%%J. P.~Blaizot, and E.~Iancu, Phys. Rep. {\bf 359}, 355 (2002).
%%%%\doi{https://doi.org/10.1016/S0370-1573(01)00061-8}

%%%%\bibitem{2}
%%%%M.~Gyulassy, and L.~McLerran, Nucl.\ Phys.\  A {\bf 750}, 30 (2005). \\ \doi{https://doi.org/10.1016/j.nuclphysa.2004.10.034}

%%%%\bibitem{3}
%%%%S.~Aoki et al. [JLQCD Collaboration], Phys. Rev. D 72, 054510 (2005). \\
%%%%\doi{https://doi.org/10.1103/PhysRevD.72.05451}

%%%%\bibitem{4}
%%%%S.~Alekhin, A.~Djouadi, and S.~Moch, Phys. Lett. B 716, 214 (2012) [arXiv:1207.0980 [hep-ph]]. %%%%\doi{https://doi.org/10.1016/j.physletb.2012.08.024}

%%%%\end{thebibliography}

%%%%\appendix

\end{document}